\documentclass[12pt,preprint]{aastex}

\usepackage{emulateapj5}
\usepackage{epsfig}
\usepackage{apjfonts}

\makeatletter

\newenvironment{figurehere}
  {\def\@captype{figure}}
  {}
\makeatother

\newcommand{\mdot}{M$_{\odot}$}
\newcommand{\edot}{$\dot{E}$}
\newcommand{\hst}{{\it HST}}
\newcommand{\cha}{{\it Chandra}}
\newcommand{\teff}{$\mathrm{T_{eff}}$}
\newcommand{\lx}{$\mathrm{L_{x}}$}
\newcommand{\wopt}{$\mathrm{W29_{opt}}$}
\newcommand{\other}{$\mathrm{W34_{opt}}$}
\newcommand{\phb}{$\phi_{b}$}

\slugcomment{Accepted for publication by {\it The Astrophysical Journal}}

\shorttitle{An MSP Optical Counterpart in 47 Tuc}
\shortauthors{Edmonds et al.}

\begin{document}

\title{A Millisecond Pulsar Optical Counterpart with Large-Amplitude
Variability in the Globular Cluster 47 Tucanae \altaffilmark{1}}

\altaffiltext{1}{Based on observations with the NASA/ESA {\em Hubble Space
Telescope} obtained at STScI, which is operated by AURA, Inc. under NASA
contract NAS 5-26555.}

\author{Peter D. Edmonds\altaffilmark{2}, Ronald
L. Gilliland\altaffilmark{3}, Fernando Camilo\altaffilmark{4}, 
Craig O. Heinke\altaffilmark{2} and Jonathan E. Grindlay\altaffilmark{2}} 
\altaffiltext{2}{Harvard-Smithsonian Center
for Astrophysics, 60 Garden St, Cambridge, MA 02138;
pedmonds@cfa.harvard.edu; cheinke@cfa.harvard.edu; josh@cfa.harvard.edu}
\altaffiltext{3}{Space Telescope Science Institute, 3700 San Martin Drive,
Baltimore, MD 21218; gillil@stsci.edu} \altaffiltext{4}{Columbia
Astrophysics Laboratory, Columbia University, 550 West 120th Street, New
York, NY 10027; fernando@astro.columbia.edu}

\begin{abstract}

Using extensive \hst\ imaging, combined with \cha\ X-ray and Parkes radio
data, we have detected the optical binary companion to a second millisecond
pulsar (MSP) in the globular cluster 47 Tucanae. This faint ($V=22.3$) blue
($V-I=0.7$) star shows a large amplitude (60--70\%) sinusoidal variation in
both $V$ and $I$. The period (3.19066 hr) and phase of the variation match
those of the MSP 47 Tuc W (which does not have an accurate radio timing
position) to within 0.5 seconds and 1.2 minutes respectively, well within
the 1~$\sigma$ errors. The phase dependence of the intensity and color
implies that heating of a tidally locked companion causes the observed
variations. The eclipsing nature of this MSP in the radio, combined with
the relatively large companion mass ($>$ 0.13 \mdot) and the companion's
position in the color-magnitude diagram, suggest that the companion is a
main sequence star, a rare circumstance for an MSP companion. This system
is likely to have had a complex evolution and represents an interesting
case study in MSP irradiation of a close companion.  We present evidence
for another optical variable with similar properties to the companion of 47
Tuc W. This variable may also be an MSP companion, although no radio
counterpart has yet been detected.

\end{abstract}

\keywords{binaries: general --- globular clusters: individual (47 Tucanae)
--- techniques: photometric --- pulsars: individual (PSR~J0024$-$7204W) ---
pulsars: general}

\section{Introduction}

More than 50 millisecond pulsars (MSPs), half of all known, have been found
in radio surveys of globular clusters, where the high stellar densities and
interaction rates cause large numbers of neutron stars (NSs) to be spun up
to millisecond periods (via binary production and subsequent accretion onto
the compact object). Studies of these objects therefore offer insight into
the formation and evolution of NS binaries (see, e.g., Rasio, Pfahl, \&
Rappaport 2000) and the frequency of stellar interactions in the dense
cores of clusters, sampling different evolutionary channels compared to the
disk population.

Recent improvements in the sensitivity of the Parkes radio telescope have
resulted in a dramatic increase in the number of cluster MSPs detected. In
particular, the 11 known MSPs in the cluster 47 Tuc (Manchester et
al. 1991; Robinson et al. 1995) have recently been increased to 20 (Camilo
et al. 2000, hereafter CLF00), with 16 of these currently having accurate
(milli-arcsecond) timing positions (Freire et al. 2001a; Freire 2001).
These enable studies of the cluster's gravitational potential well (Freire
et al. 2001a) and of the intra-cluster medium (Freire et al. 2001b).

Another significant observational advance has been the ability of \cha\ to
detect cluster MSPs in large numbers (Grindlay et al. 2001a, 2002; D'Amico
et al. 2002).  Accurate (0\farcs1--0\farcs2) X-ray positions for the MSPs
combined with the detection of large numbers of cataclysmic variables (CVs)
and active binaries with both \cha\ and \hst\ (Grindlay et al. 2001a and
Edmonds et al. 2002a, in preparation) has allowed the radio, X-ray and
optical data to be placed on a common astrometric frame, good to
$\lesssim$0\farcs1.

These advances, and the superb spatial resolution of \cha\ and \hst, have
allowed the recent detection of two optical counterparts to cluster MSPs,
PSR~J0024$-$7203U, hereafter 47 Tuc U, in 47 Tuc (Edmonds et al. 2001a,
hereafter EGH01), and PSR~J1740$-$5340, hereafter NGC 6397 A, in NGC 6397
(Ferraro et al. 2001). The MSP 47 Tuc U has a $\sim$0.15 \mdot\ He white
dwarf (WD) companion, a 10.3 hr binary period and small amplitude (0.004
mag in $V$) orbital variations caused by heating of one side of the
companion (EGH01).  The $\sim$0.45 \mdot\ companion in NGC 6397 A is either
a subgiant or a heated main sequence (MS) star (Ferraro et al. 2001). The
32.5 hr, 0.1 mag orbital variations are caused by tidal distortion of the
secondary by the NS, and are roughly sinusoidal with a (32.5/2) hr
period. Another notable feature of NGC 6397 A is that the spectrum of the
X-ray counterpart is relatively hard, and is suggestive of non-thermal
emission. By contrast, 8 out of 9 of the 47 Tuc MSPs which are bright
enough to have useful spectral information are soft (Grindlay et al. 2002).
 
Currently, NGC 6397 A appears to be unique among known MSPs in having a
likely non-degenerate companion. The MSPs in 47 Tuc, for example, are
either single, have likely He WD companions (e.g. 47 Tuc U), or have
secondaries with masses of $\sim$0.02--0.03 \mdot\ (CLF00) which cannot be
normal MS stars.  As argued by Burderi et al. (2002) and Ferraro et
al. (2001), the subgiant companion to NGC 6397 A may have been the star
that recently spun up the MSP, consistent with the relatively small
(characteristic) age for the system (D'Amico et al. 2001). The position of
the star in the color-magnitude diagram (CMD) is roughly consistent with
that of a mass-losing subgiant like that in the CV AKO~9 in 47 Tuc (Albrow
et al. 2001).  The equally interesting possibility presented by Ferraro et
al. (2001) is that an MS star has been captured in an exchange interaction
with an NS already spun-up to millisecond periods. Further work is needed
to test whether the position of the star in the CMD is consistent with
estimates of heating of the companion by the MSP (Ferraro et al. 2001).
The relatively large amplitude of the ellipsoidal variations in NGC 6397 A
suggests a high inclination orbit (Ferraro et al. 2001; Orosz \& van
Kerkwijk 2002), consistent with the detection of eclipses in the radio
(D'Amico et al. 2001).

Five out of eight eclipsing MSPs known have very low companion masses of
$\sim$0.02--0.03 \mdot, and short orbital periods (these include the 47 Tuc
MSPs J, O, and R). The three exceptions are NGC 6397 A; PSR~B1744$-$24A in the
cluster Terzan~5, with an orbital period of $\sim$1.8 hr, companion mass
$\sim$0.10 \mdot, and displaying irregular eclipses (Lyne et al. 1990); and
47 Tuc W, an eclipsing MSP in 47 Tuc with a $\sim$3.2 hr period but with a
companion mass of $\sim$0.15 \mdot\ (CLF00; see \S~\ref{sec.radio}).

We present here extensive \hst\ evidence for an optical counterpart to 47
Tuc W based on the detection of a faint, large amplitude, optical variable
with an orbital period and phase which perfectly match those of 47 Tuc W
within the uncertainties. The X-ray counterpart detected with \cha\ has
properties which are similar to those of NGC 6397 A. This detection
represents the best evidence yet obtained for an MSP with an MS companion,
given uncertainties about the evolutionary status of NGC 6397 A.  Following
a brief summary of the radio data available on 47 Tuc W, we present the
astrometry, absolute photometry and time series photometry for the 47 Tuc W
optical companion in \S~2, where we also present evidence for a 2nd
possible MSP companion with similar optical and X-ray properties. In \S~3
we discuss the implications of these results.

\section{Observations, data analysis, and results}

\subsection{Radio Data}\label{sec.radio}

The 2.35\,ms pulsar 47 Tuc W was discovered on 1999 February 5 (MJD 51214)
during a $\sim$4 hr observation with the Parkes radio telescope (CLF00).
It has never been detected again (which happens for a few very weak pulsars
that are detected seldom and only due to the focusing effects of
interstellar scintillation), and hence has no position determined from
radio timing measurements.  However, that one observation was sufficient to
determine that it is part of a binary system with a period of $\sim$3.2 hr,
owing to the varying apparent spin period caused by Doppler shift.  The
times-of-arrival (TOAs) determined from the discovery data can be used to
obtain precise orbital parameters.

We have used 40 good TOAs together with the {\sc tempo}\footnote{See
http://pulsar.princeton.edu/tempo.} timing software to measure an accurate
barycentric pulsar spin period, projected semi-major axis of $0.2434(1)$
light-sec, binary period $P_b = 0.13295(4)$ days, and time of pulsar
ascending node $T_0 = 51214.284829(3)$ MJD, where all uncertainties given
are conservative ($10\times$ {\sc tempo}), accounting for covariances in
fitted parameters.  The orbit is circular, with eccentricity $<0.001$.
With orbital phase (\phb) zero defined at $T_0$, the pulsar is clearly
eclipsed during 0.1$<$ \phb\ $<$0.4 (hereafter we redefine the fiducial
point of orbital phase such that \phb\ = 0 at $T_{max} \equiv T_0 +
[3/4]P_b$, retaining the binary phase convention used by EGH01).  With the
measured orbital parameters, and assuming an NS mass of 1.4 \mdot, the
pulsar mass function implies a companion mass $>0.13$ \mdot, with a mass of
$\sim$0.15 \mdot\ for an orbital inclination of 60\degr.

\subsection{Summary of Optical Data}

The optical data used in this paper consists of four different \hst\
programs, GO-8267, GO-7503, GO-8219, and SM3/ACS-9028, summarized in Table
1. The program GO-8267 (PI R. Gilliland) involved an extensive set of 160~s
Wide Field Planetary Camera 2 (WFPC2) exposures in F555W (636 images) and
F814W (653) obtained over 8.3 days in 1999 July. A limited number (28) of
F336W exposures were also obtained (not shown in Table 1).  The use of
these data to search for transiting planets and to study binaries are
described in Gilliland et al. (2000) and Albrow et al. (2001) respectively.
The program GO-7503 (PI G. Meylan) involved a short set of dithered F555W
exposures with WFPC2 taken over three orbits about 110 days after GO-8267.
The GO-8219 (PI C. Knigge) data consisted of six blocks of eight 30~s
exposures each using the Space Telescope Imaging Spectrograph (STIS) in CCD
mode with the so-called CLEAR filter (two exposures of one visit were lost resulting
in 46 total exposures).  The high throughput of STIS/CCD resulted in
comparable count levels for faint stars in 30~s compared to the 160~s F555W
and F814W WFPC2 exposures.  The first visit of GO-8219 was about 65 days
after GO-8267, while the remaining five visits were over a 10 day period
some 401 days after GO-8267.  Finally, SM3/ACS-9028 consisted of 20
cosmic ray split observations of 120~s with the F475W filter using the High
Resolution Camera (HRC) of the newly installed Advanced Camera for Surveys
(ACS) on \hst. These data were acquired with 20 large-scale dithers over 4
orbits of \hst\ as calibration of the camera's geometric distortion, and
were obtained in 2002 April about 1002 days after GO-8267.

\subsection{W29 and \cha--\hst\ Astrometry}
\label{sec.ast}

The astrometry between \cha\ and \hst, based on the GO-8267 data, has
already been described in EGH01 and Edmonds et al. (2002b). A partial set
of optical identifications was described by Grindlay et al. (2001a) and the
full set is described in Edmonds et al. (2002a, in preparation).  This
paper focuses on the optical identification for the X-ray source W29 from
Grindlay et al. (2001a). We first establish the similarity of this source
to some known X-ray counterparts of MSPs, and then present the \cha--\hst\
astrometry.

The low luminosity and relatively hard X-ray spectrum of W29 are similar to
those of U12 (the X-ray counterpart of NGC 6397 A; Grindlay et al. 2001b) and
47 Tuc J, both eclipsing MSP binaries (see Grindlay et al. 2002).  The
X-ray spectrum of W29 can be best fit with a power law (photon index =
1.8$\pm$0.6 when fixing $N_H$ to the nominal cluster value; $\chi^2_{\nu} =
1.55$) and the luminosity in the 0.5--2.5 keV band is $7.8
\times10^{30}$erg s$^{-1}$.  By comparison, for U12 a power law fit gives a
photon index of 1.6$\pm$0.3 (when fixing $N_H$ to the cluster value) and a
luminosity of $4 \times10^{30}$erg s$^{-1}$. Because both W29 and U12 are
relatively faint sources (41 and 66 counts respectively), these X-ray
sources are also consistent with $\sim$6~keV thermal bremsstrahlung spectra
($\chi^2_{\nu} = 1.94$ for W29). The X-ray counterpart of 47 Tuc J, W63,
has only 10 counts (\lx\ = $1.8 \times10^{30}$erg s$^{-1}$) and is too
faint for useful spectral fitting, but has hardness ratios consistent with
a power law of photon index between 1 and 1.5 (Grindlay et al. 2002).

\begin{figurehere}
\vspace*{0.3cm}
\hspace*{-0.2cm}
\epsfig{file=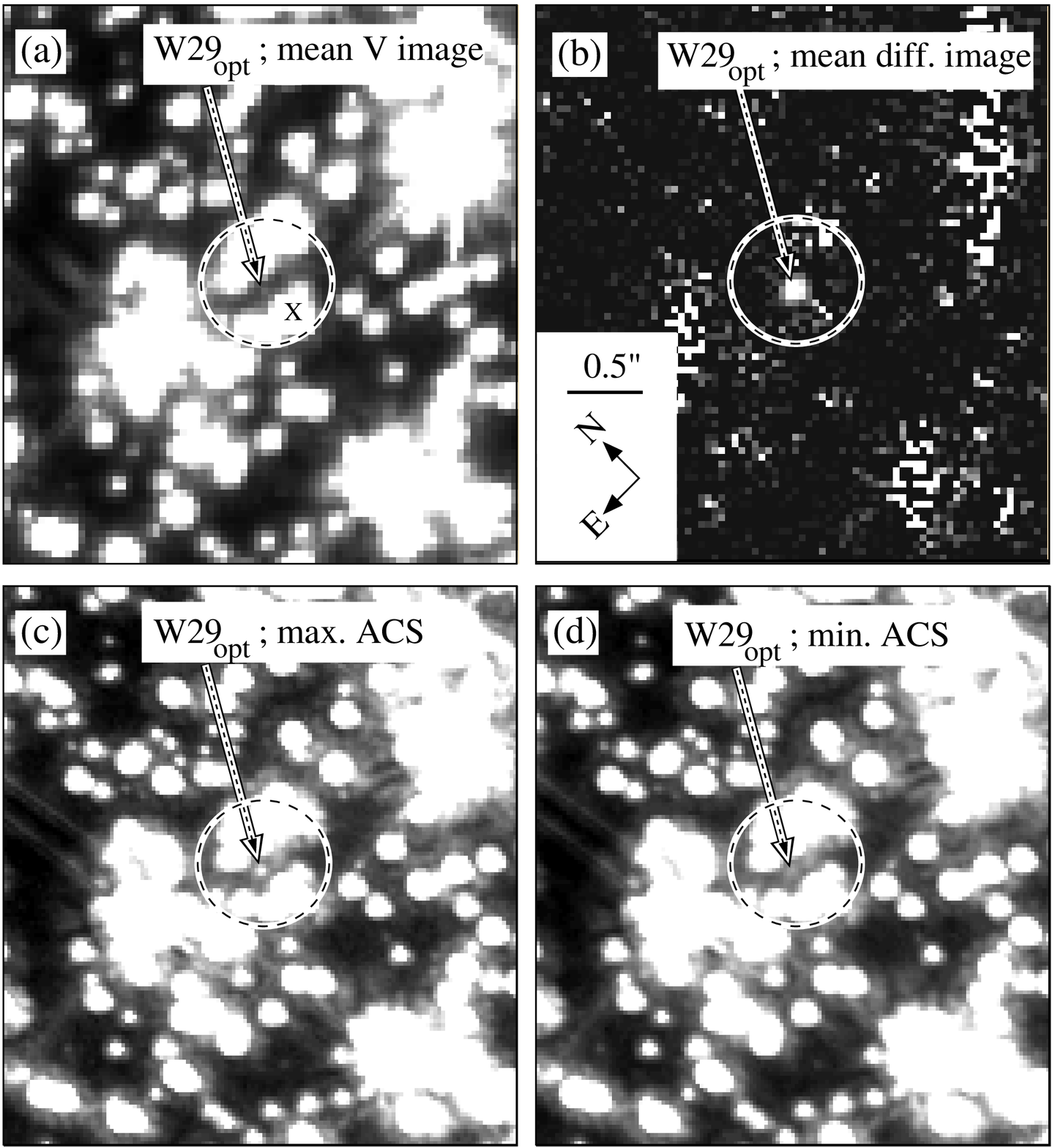,width=8.5cm}
\vspace*{0.2cm}
\caption{\hst\ finding charts for \wopt, using the GO-8267 ($V$) and
SM3/ACS-9028 (F475W) data. The mean direct image from GO-8267 (WFPC2) is
shown in Fig. \ref{fig.fchart}a and the mean difference image is shown in
Fig. \ref{fig.fchart}b (see text for definitions). The ACS/HRC image for
\wopt\ near maximum (\phb\ = 0.0$\pm$0.16) is shown in
Fig. \ref{fig.fchart}c and the corresponding image near minimum (\phb\ =
0.5$\pm$0.16) in Fig. \ref{fig.fchart}d.  The 5~$\sigma$ (0\farcs42) error
circle for W29 is shown, as is the scale and orientation of the \hst\
images. The variable, optical counterpart for W29 (denoted \wopt) is
clearly visible near the center of the X-ray error circle in the mean
difference and maximum ACS images (labeled with an arrow), but is barely
visible in the mean direct and minimum ACS images. The initial
identification for the $\sim$3.2 hr variable (PC1-V46) is labeled with
`X'.}
\vspace*{0.3cm}
\label{fig.fchart}
\end{figurehere}

A plausible optical counterpart for W29 was identified as a nearby 3.187 hr
period variable. The variations in this very crowded region were initially
believed to arise from a relatively bright star ($U = 18.37$, $V = 17.79$
and $I = 17.13$), identified as a non-eclipsing contact binary with an
orbital period of 6.374 hr, and an amplitude of $\sim$0.002 mag (PC1-V46
from Albrow et al. 2001).  This star is located 0\farcs25 (3.0~$\sigma$)
from W29, where the astrometric uncertainty standard deviation $\sigma$ is
a combination of positional errors for W29 and small systematic errors in
transforming between the \cha\ and \hst\ coordinate frames.  However, a
careful examination of the directly observed images (`direct images'; a
term used to describe both individual exposures and mean images) and
difference images\footnote{difference images are generated by subtracting
from each individual direct image the mean over-sampled image evaluated at
the appropriate dither position; see Albrow et al. 2001} showed that the
bright star identified as PC1-V46 was not the true variable. Instead, there
is a faint, nearby star (only 0\farcs19 away) with a much larger intrinsic
variability amplitude.  This star was not included in our original star
lists because it is faint and very crowded (it lies only 3\farcs8, or 16\%
of the core radius, $r_c$, away from the cluster center), but it fell
within, and hence contaminated, the aperture used for PC1-V46. The new
variable will be designated \wopt\ since it is only 0\farcs07
(0.8~$\sigma$) away from W29.  The likelihood that this positional match is
a coincidence is very low. On the PC chip there are 39 variables from
Albrow et al. (2001) with $V$ fainter than the MS turnoff, so the
probability of a chance match between W29 and an optical variable is only
$\sim 5\times 10^{-4}$.  The position of \wopt\ in the astrometric
coordinate system of the MSPs (based on the JPL DE200 planetary ephemeris;
Freire et al. 2001a) is given in Table 2.

The time series for \wopt\ is approximately sinusoidal with an amplitude of
70\% in the $V$-band and 60\% in the $I$-band. More details on the time
series will be given in \S~\ref{sec.tseries}. Here, we use the phase
information for \wopt\ to construct mean direct and difference images for
this variable.  Figure \ref{fig.fchart}a shows the average of direct images
within $\pm$0.07 in orbital phase of \phb\ = 0.0 (maximum light) and within
$\pm$0.07 of \phb\ = 0.5 (minimum light) in the \wopt\ time series, where
we use the binary phase convention given in \S~\ref{sec.radio}. This provides an
average intensity direct image for the variable without over-sampling.  The
variable \wopt\ is barely visible about 1 pixel below the center of the
5~$\sigma$ X-ray error circle for W29.  Clearly, \wopt\ is located in a
crowded region of 47 Tuc even at \hst\ resolution.  Although there are no
other identifiable stars within a 0\farcs1 radius of \wopt, within a radius
of 0\farcs2 there are three stars with a total intensity in F555W over 200
times that of \wopt, and within a 0\farcs5 radius a total of 13 stars have
a total intensity nearly 800 times that of \wopt.  Since \wopt\ has a color
comparable to cluster turnoff stars (see \S~\ref{sec.phot}), while most
stars are redder than this, the relative brightness of nearby stars is
somewhat higher yet in F814W or the STIS CLEAR filter bandpass.

Figure \ref{fig.fchart}b shows the sum of all difference images within
$\pm$0.07 of \phb\ = 0.0, minus the sum of all difference images within
$\pm$0.07 of \phb\ = 0.5, giving the mean difference image.  Non-variables
disappear except for noise. Note the clear detection of \wopt\ near the
center of the X-ray error circle and coincident with a faint, relatively
blue star just visible in the mean direct image. The large variability of
\wopt\ is shown by comparing the two ACS/HRC images shown near maximum
(\phb\ = 0.0$\pm$0.16) and minimum (\phb\ = 0.5$\pm$0.16).  The absolute
photometry for this star will now be described.

\subsection{Photometry}
\label{sec.phot}

Because of the extreme crowding described in \S~\ref{sec.ast}, the absolute
photometry for \wopt\ requires special attention and we have used two
different methods to estimate the $U$, $V$ and $I$ magnitudes for this
star:

({\em i}) We constructed direct images for \wopt\ at maximum (\phb\ =
0.0$\pm$0.16) and minimum (\phb\ = 0.5$\pm$0.16) in $V$ and $I$, and
subtracted the minimum image away from both the maximum and the mean
images. In each of these two residual images, \wopt\ was much more clearly
visible than in any of the direct images, and crowding was no longer a
limiting factor in the photometry. After carrying out point spread function
(PSF) fitting to the \wopt\ neighbors in the mean image, we applied this
PSF-fitting to \wopt\ in the residual images, and used the measured light
curves to derive $V$ and $I$ magnitudes for \wopt\ at \phb\ = 0.0, \phb\ =
0.5, and mean values averaged over the complete GO-8267 dataset. For $U$ a
mean value only was derived using this technique, because of the much lower
signal-to-noise ratio. Mean values of $U = 23.8 \pm 0.5$, $V = 22.3 \pm
0.2$ and $I = 21.6 \pm 0.3$ were derived (see Table 2), where the errors
are a combination of uncertainties from the PSF-fitting and estimated
absolute errors in the conversion from \hst\ to Johnson magnitudes.

({\em ii}) The second technique involved use of the deep, oversampled
images, the accurately known position for \wopt, and local sky
determinations, to estimate the peak intensity levels for \wopt\ in the
three different filters. We then scaled these values to Johnson $UVI$
magnitudes using nearby stars and the PSF-fitting results of Gilliland et
al. (2000). Mean values of $U = 23.7 \pm 0.7$, $V = 22.4 \pm 0.2$ and $I =
21.7 \pm 0.3$ were derived, in good agreement with the values from ({\em
i}).

\begin{figurehere}
\vspace*{-0.3cm}
\hspace*{-0.5cm}
\epsfig{file=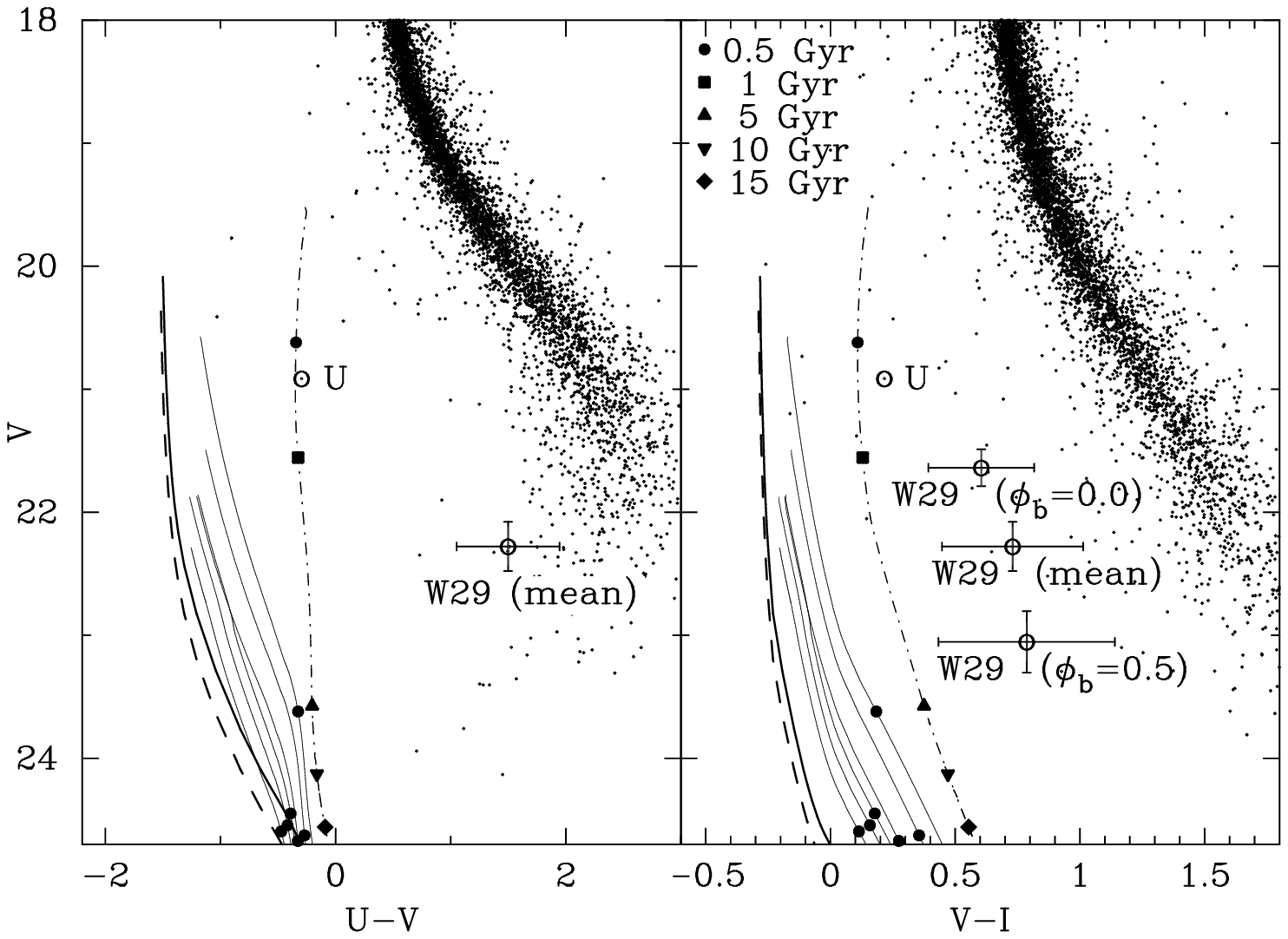,width=9.0cm}
\vspace*{-0.2cm}
\caption{\hst\ CMDs for GO-8267 based on PSF-fitting. The variable \wopt\
and the counterpart to the MSP 47 Tuc U (`U'; EGH01) are labeled. The
thicker lines show \citet{ber95} models for 0.5 \mdot\ (dashed line) and
0.6 \mdot\ (solid line) CO WDs. The thinner lines are \citet{ser01} models,
for He WD masses of 0.406, 0.360, 0.327, 0.292, 0.242, 0.196 \mdot, with
mass decreasing towards the red, and 0.169 \mdot\ (dot-dashed
line). Cooling ages are as shown.  These models have been plotted assuming
that $(m-M)_0=13.27$, the mean of the 9 distance modulus values reported in
\citet{zoc01}, $E(B-V)=0.055$ \citep{zoc01}, $A_U/A_V=1.51$ and
$A_I/A_V=0.60$ \citep{hol95}. Using $(m-M)_0=13.27$ the distance to 47 Tuc
is 4.5 kpc.}
\vspace*{0.6cm}
\label{fig.cmds}
\end{figurehere}

Figure \ref{fig.cmds} shows the $V$ vs $U-V$ and $V$ vs $V-I$ CMDs for the
PC chip, with \wopt\ labeled. The known counterpart to MSP 47 Tuc U (EGH01)
is also shown. All of the points in the CMDs, including \wopt\ (where the
photometry from [{\em i}] was used), had magnitudes which were
self-consistently determined by PSF fitting. The various plotted lines are
the CO WD cooling tracks of Bergeron et al. (1995) and the He WD cooling
tracks of Serenelli et al. (2001). Note that the colors of \wopt\ appear to
be inconsistent with those of any He WD cooling track, unlike the case for
47 Tuc U.

\subsection{Time Series}
\label{sec.tseries}

\subsubsection{GO-8267 data}

The original time series analysis, upon which the initial possible match of
\wopt\ with the MSP 47 Tuc W was made, used only the GO-8267 data.  Here,
we applied the standard difference image technique (see Gilliland et
al. 2000 and Albrow et al. 2001) using a circular aperture, for this
crowded target star, with an area of only 5 pixels (1 pixel =
$0\farcs045\times 0\farcs045$) to minimize contamination from neighboring
companions.  The top panel of Figure \ref{fig.w29tseries} shows the full
time series for \wopt, and the middle and bottom panels show the GO-8267
data with a sinusoidal fit superimposed (the $I$-band light curve has been
scaled by a factor of 1.17 so that $V$ and $I$ have the same amplitude).
The times displayed in the x-axis were derived by extracting the Modified
Julian Date (MJD) of the observations, given in Universal Time, from the
\hst\ FITS headers, and converting to Heliocentric Julian Date (HJD) using
the IRAF task {\tt rvcorrect}\footnote{HJD differs from barycentric MJD by
a variable heliocenter--barycenter light travel time of $\sim$1--2~s, in
addition to a trivial 0.5 d offset due to a different zero-point definition
of the two time scales.}.  The y-axis units are fractional intensity,
defined as (intensity$-<$intensity$>$)/$<$intensity$>$ (where
$<$intensity$>$ is the mean intensity). This system is used because it is
linear (unlike the magnitude system) and gives light curves for \wopt\
which correctly appear sinusoidal for an intrinsic sinusoidal modulation.

\begin{figurehere}
\vspace*{0.3cm}
\hspace*{-0.2cm}
\epsfig{file=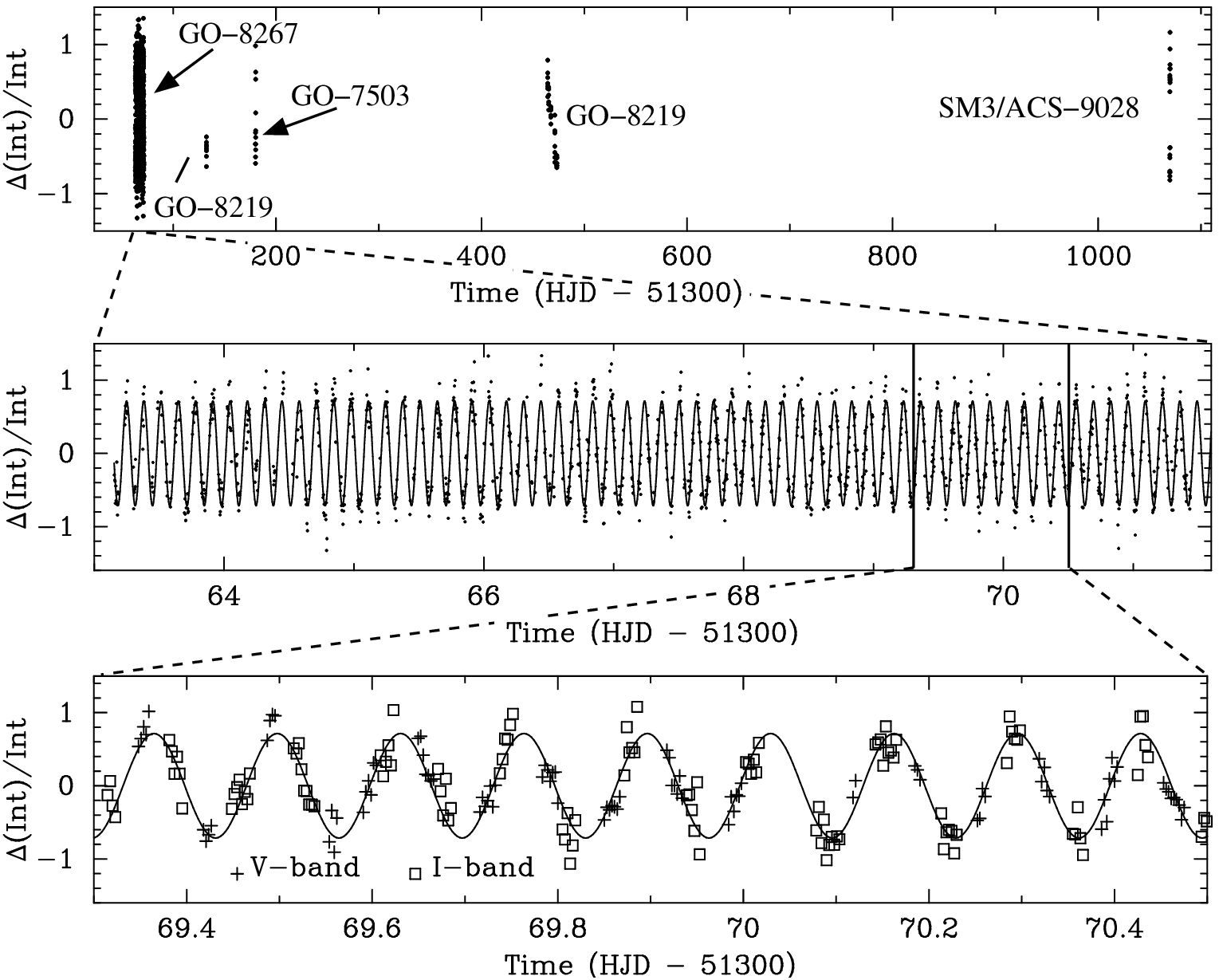,width=8.5cm}
\vspace*{0.1cm}
\caption{\hst\ time series for \wopt. The top panel shows the full set of
time series analyzed here (the different datasets are labeled), extending
from 1999 July to 2002 April. The times shown are extracted from the \hst\
FITS headers.  The middle panel shows the GO-8267 data and the bottom panel
shows a close-up of a typical sample of the GO-8267 data (a sinusoidal fit
to the data is shown in each case). The $I$-band data has been scaled so
that its amplitude equals that of the $V$-band data. In the bottom panel
different symbols have been used for the two filters.}
\vspace*{0.5cm}
\label{fig.w29tseries}
\end{figurehere}

We then performed a simultaneous least-squares fit of the GO-8267 $V$ and
$I$ data to a sinusoid, allowing the amplitude, period and time of optical
maximum ($T_{max} \equiv T$[\phb\ = 0.0]) to vary. The following values
were obtained: amplitude = 0.714(9), $P_b = 0.132919(16)$ d, and $T_{max} =
51366.87064(28)$ MJD (all $T_{max}$ values are given in barycentric MJD to
within a small error of 1--2~s). Because of the difficult photometry for
\wopt\ there is an additional $\sim$20\% error for the amplitude resulting
from assumed normalization using the mean absolute magnitude.

We now compare these results with the measured radio parameters of 47 Tuc
W.  The orbital period of 47 Tuc W (see \S~\ref{sec.radio} and Table 2)
differs from the period of \wopt\ by only 0.00003 d (2.7~s, a 0.02\%
relative difference), a difference of 0.8~$\sigma$ when using the larger (radio)
error. The radio value for $T_{max}$ (when the heated hemisphere of \wopt\
is facing towards the Sun and is expected to be brightest, in a model where
\wopt\ is heated by the pulsar wind) is 51214.38454(3) barycentric MJD
(Table 2). By projecting the optically-derived $T_{max}$ backwards in time
by 1147 orbits, we derive $T_{max}$ = 51214.412(18) MJD. This differs from
the radio-inferred $T_{max}$ by (0.028 $\pm$ 0.018) d, a 40 minute
difference (0.2 in phase). These results suggest that \wopt\ may be the
companion to 47 Tuc W, although the phase errors are still relatively
large. These errors are significantly reduced in \S~\ref{sec.archive} with
the use of archival data.

\begin{figure*}
\vspace*{0.3cm}
\hspace*{2.0cm}
\epsfig{file=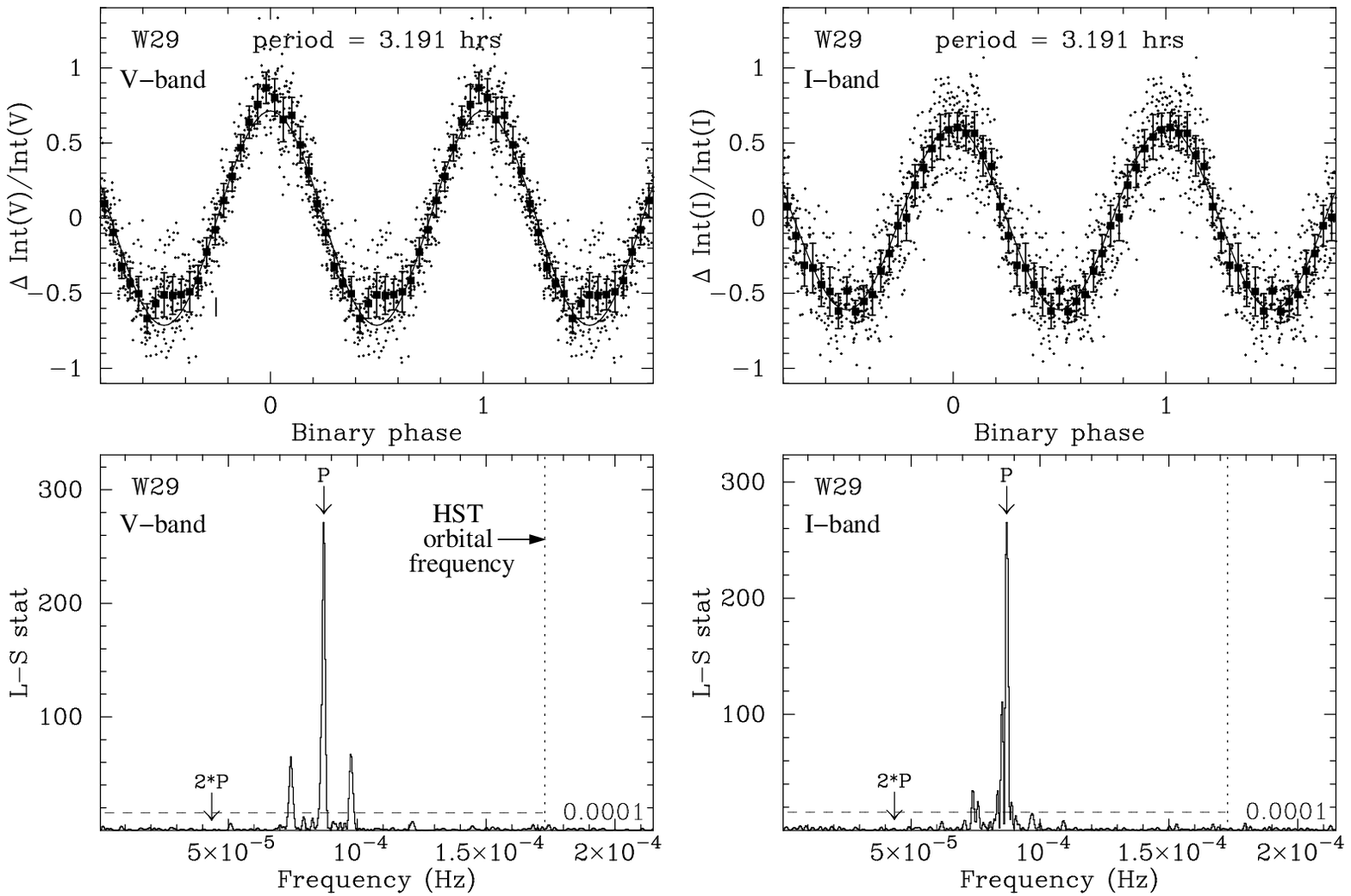,width=13.5cm}
\vspace*{0.0cm}
\caption{Phase plots and power spectra for \wopt, using the GO-8267
data. The left side shows the $V$-band data and the right side the $I$-band
data. The period used to generate the phase plots is slightly (0.02\%)
higher than the value derived from the GO-8267 data alone, because of
improvements based on archival analysis (described in
\S~\ref{sec.archive}). The squares show the mean fractional intensity in 25
phase bins between \phb\ = 0.0 and \phb\ = 1.0, where the error bars give
the 3~$\sigma$ errors ($\sigma$ is the standard error in the mean). The
solid line shows a sinusoid with the fitted amplitude, period and
phase. Note the differences between the $V$-band data and the sinusoidal
model near \phb\ = 0.0 and \phb\ = 0.5 (see \S~\ref{sec.colchange}). The
vertical dotted line in the power spectrum shows the \hst\ orbital period
(96.4 min) and the horizontal dashed line shows the Lomb-Scargle power
corresponding to a false-alarm probability (Scargle 1982) of 1 $\times
10^{-4}$. The power spectrum peaks near the main period are aliases caused
by the \hst\ window function.  }
\vspace*{-0.4cm}
\label{fig.w29phase}
\end{figure*}

The phase plots and Lomb-Scargle (Scargle 1982) power spectra for \wopt\ in
the $V$ and $I$ bands are shown in Figure \ref{fig.w29phase}. Note the
similarity of the time series (for both filters) to a sinusoid. The power
spectrum shows a large peak at the frequency value corresponding to the
fitted period, with no evidence for a peak at twice the fitted period.  The
modulation is thus consistent with arising from a single heated face of the
secondary in a circular orbit.

\subsubsection{Use of archival data}
\label{sec.archive}

With the above solution as a starting point, we used images from the
archive (GO-7503, GO-8219 and SM3/ACS-9028) to refine the orbital solution
for the period and phase of \wopt. The errors for the period and phase of
\wopt\ from GO-8267 are small enough that the archival data can be included
in the orbital solution without any cycle count ambiguity. For example, the
401 day separation between GO-8267 and GO-8219 gives a phase uncertainty of
0.048 d (0.36 of one orbit), when using the error quoted above for the
period. Therefore, there is potential to significantly improve our orbital
solution using these extra data.

The small set of F555W data from GO-7503 was taken in one contiguous period
and spans about 0.85 in phase for \wopt.  The combination of multi-pixel
dithers for the small number of exposures from GO-7503, coupled with the
strong variability of \wopt, implied that the GO-7503 data itself was not
sufficient to generate difference images relative to an internally
generated over-sampled mean image.  To obtain difference images necessary
for extracting variations of \wopt\ from the GO-7503 data we used the mean,
over-sampled image from GO-8267.  Positions were fitted to about 50 stars
within 7$''$ of \wopt\ in both the over-sampled reference image and each
individual exposure from GO-7503.  Transformation coefficients were derived
including x,y offsets, rotation and independent plate scales in x,y from
the two sets of star positions.  Then cubic-spline interpolation in the
(essentially noiseless) over-sampled reference image from GO-8267 was used
to evaluate an image at the position corresponding to each GO-7503
exposure.  Difference images were then formed and the photometry at \wopt\
was obtained as a 5 pixel sum in each difference image at the calculated
position of \wopt.  Since this is a very crowded region, and the two data
sets had an orientation difference of 113\degr, there was a net (arbitrary)
zero point offset removed to force consistency between the relative
photometry of the GO-7503 and GO-8267 datasets.  The same scale factor
(mean intensity of \wopt\ in the direct image) was used for GO-7503
(compensating for relative exposure times) as for GO-8267, taken with the
same filter and detector.

As with the GO-7503 data, there were not enough independent dithers and
exposures for GO-8219 to generate good difference images from these data
alone.  The GO-8219 data were rotated by $\sim$54\degr\ relative to
GO-8267, and were taken with a different instrument/filter combination.
Again, fits were performed for about 50 stars within 7$''$ of \wopt\ in
both the PC1 over-sampled reference image and the individual STIS
exposures.  Since the STIS CLEAR filter has high response spanning the
WFPC2 F555W and F814W filters, we evaluated reference images using cubic
spline interpolation from each of these two over-sampled images.  The match
to STIS data was then evaluated as a linear combination of F555W and F814W
reference images and zero point offset, providing a best fit to 50 fiducial
stars and several nominal regions representing sky.  A mean contribution of
0.437 times $V$ and 0.590 times the $I$ images from PC1 were adopted to
best represent the STIS CLEAR data and difference images were formed.  To
further match the STIS and WFPC2 data images we evaluated and applied
convolution kernels attempting to match the relative PSFs.  A best solution
was found using a convolution kernel that broadened individual STIS PSFs by
10--15\% relative to those for the PC1 references.  The new difference
images were then used in a standard way to extract relative intensity
values for \wopt\ using 5 pixel sums at the calculated position of \wopt\
in each frame.  Here, relative to the WFPC2 data sets, there is an
additional uncertainty in normalization of the relative photometry for
\wopt\ at the $\sim$25\% level.

\begin{figurehere}
\vspace*{0.5cm}
\hspace*{-0.2cm}
\epsfig{file=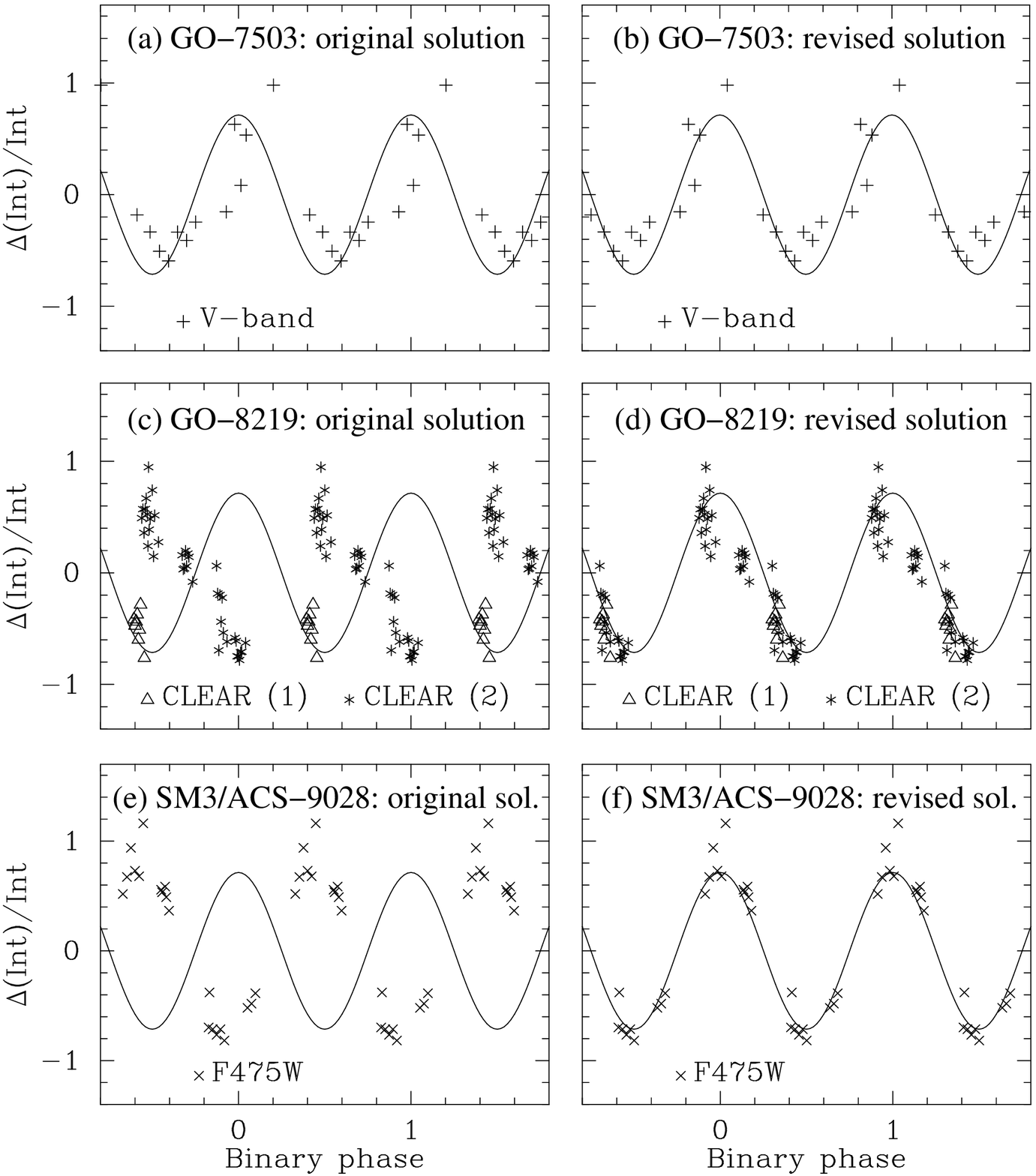,width=8.5cm}
\vspace*{0.1cm}
\caption{Phase plots for archival data from GO-7503, GO-8219 and
SM3/ACS-9028. Figure \ref{fig.arphase}a shows the sinusoid derived
from the GO-8267 data only (the `original solution') applied to GO-7503,
and Figure \ref{fig.arphase}b shows the revised solution based
on the use of all four datasets. Similarly, Figures \ref{fig.arphase}c and
\ref{fig.arphase}d show the original and revised solution for GO-8219
(where `CLEAR (1)' refers to the 1999 data and `CLEAR(2)' refers to the
2000 data) and Figures \ref{fig.arphase}e and \ref{fig.arphase}f the
results for SM3/ACS-9028. Note the dramatic improvement in the results for
GO-8219 and SM3/ACS-9028 based on the revised solution.}
\vspace*{0.5cm}
\label{fig.arphase}
\end{figurehere}

For SM3/ACS-9028 the improved spatial resolution and PSF for HRC/ACS
compared to WFPC2 permitted direct aperture photometry for \wopt, rather
than difference image analysis (the star is clearly visible in the direct
images). The data were dithered by close to half the detector size as part
of this calibration program, but \wopt\ was found in all of the frames.

We then applied the `original' sinusoidal solution from the 8.3 d of
GO-8267 data to the archival time series.  Plots of the GO-7503 data
($\sim$110 d after GO-8267) using the above period of 0.132919(16) d
(Fig. \ref{fig.arphase}a) showed a slight phase offset requiring a period
increase of 0.000025 d ($\sim 1.6\sigma$ using original errors).  Plots of
the GO-8219 data using the original period (Fig. \ref{fig.arphase}c) showed
that after $\sim$401 d the phase error had grown to nearly 0.6, and by the
SM3/ACS-9028 data 1002 days later, the phase error had grown to 1.4.
Fortunately, the combination of the GO-8267, GO-7503, GO-8219 and
SM3/ACS-9028 data allows all four data sets to be phased with no cycle
count ambiguity.  The resulting period and phase when fitting all four sets
of data are $P_b = 0.132944(1)$ d and $T_{max}$ = 51366.8705(3) MJD (see
Table 2).  By using the archival data the error in the period has been
reduced by a factor of 16, with no error change in $T_{max}$.

\begin{figurehere}
\vspace*{0.3cm}
\hspace*{0.2cm}
\epsfig{file=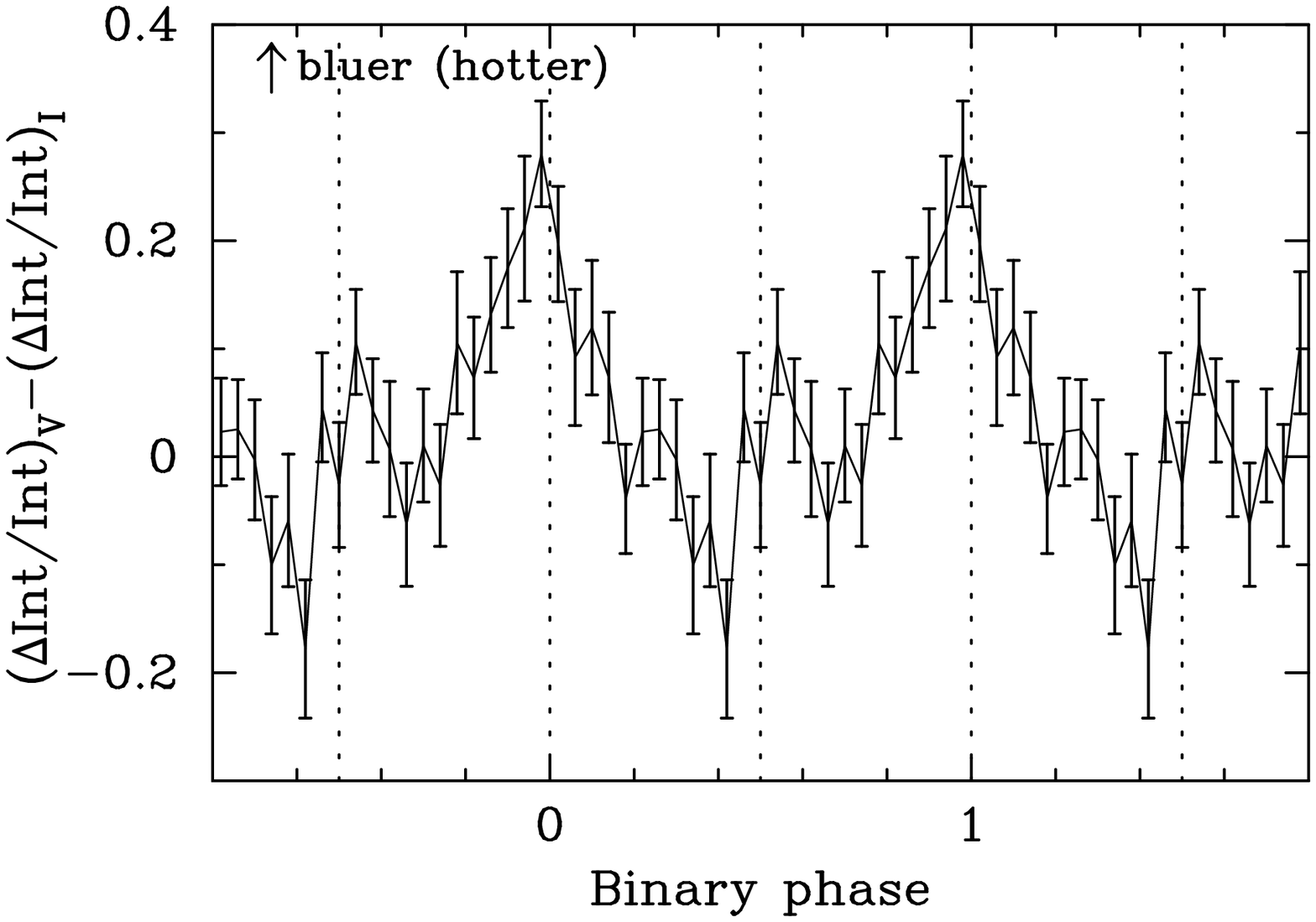,width=7.0cm}
\vspace*{0.0cm}
\caption{The difference between the $V$ and $I$ time series from GO-8267
plotted as a function of phase for \wopt. The variable is hottest when it
is brightest (\phb\ = 0.0) and coldest at \phb\ $\sim$0.4.}
\vspace*{0.5cm}
\label{fig.w29col}
\end{figurehere}

\begin{figurehere}
\vspace*{0.3cm}
\hspace*{-0.2cm}
\epsfig{file=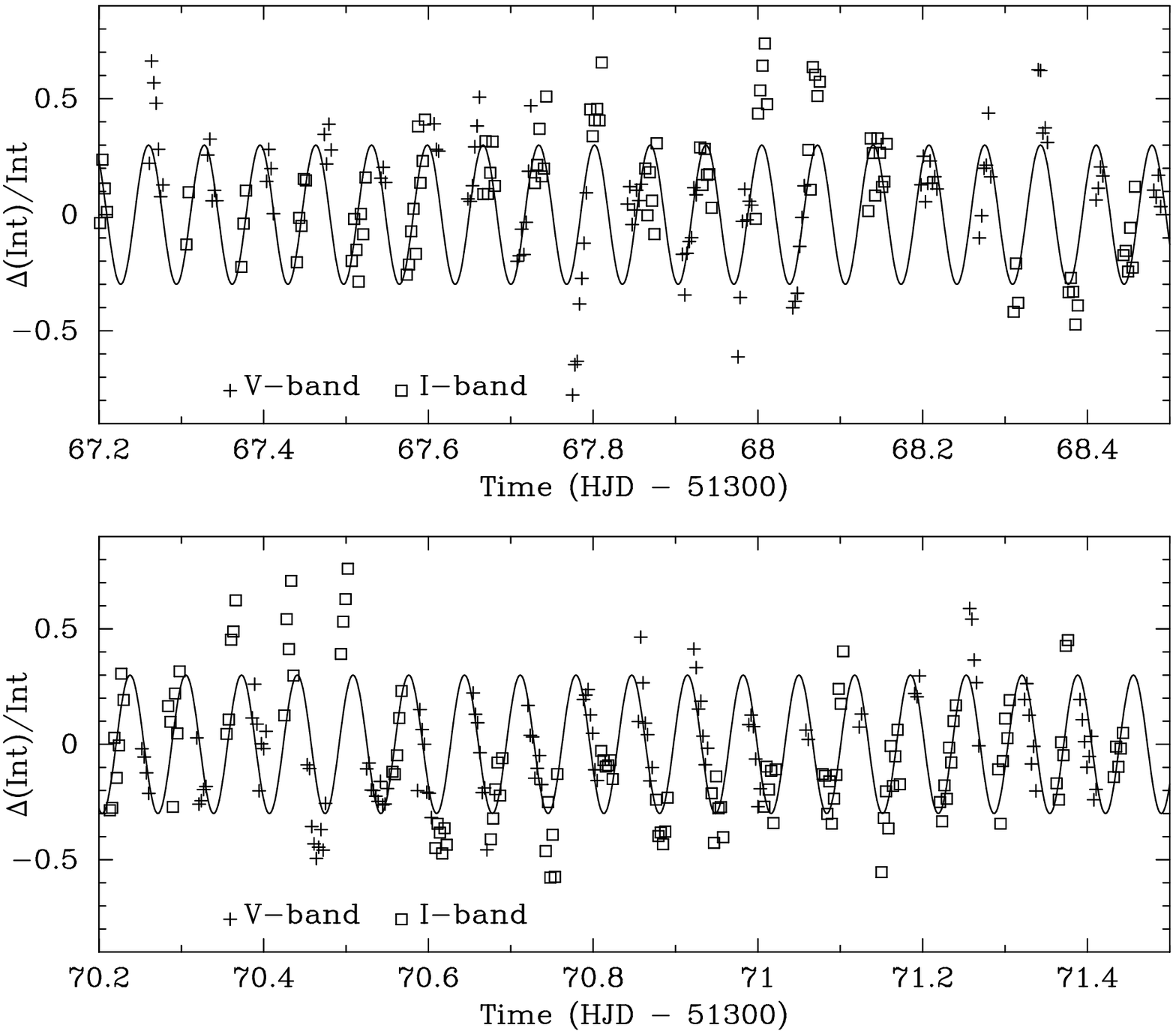,width=8.5cm}
\vspace*{0.0cm}
\caption{Segment of \other\ time series for the GO-8267 data, showing both
the $V$ and $I$ time series. A sinusoidal fit to the time series is
shown. Note the large deviations from this fit near HJD $\sim$51367.8
and HJD = 51370.35--51370.5.}
\vspace*{0.5cm}
\label{fig.w34tseries}
\end{figurehere}

Using these improved results, the periods of 47 Tuc W and \wopt\ now differ
by 0.000006 d (0.5~s), a difference of 0.2~$\sigma$ when using the larger
(radio) error. Projecting $T_{max}$ for \wopt\ backwards in time by 1147
orbits, we derive $T_{max}$ = 51214.3837(12) MJD, which differs from the
radio-inferred value by only (0.0008 $\pm$ 0.0012) d, a 1.2 minute
difference (0.006 in phase). This combination of excellent binary period
and phase agreement between \wopt\ and 47 Tuc W ensures that \wopt\ is the
optical counterpart to the MSP.

\begin{figure*}
\vspace*{0.3cm}
\hspace*{2.0cm}
\epsfig{file=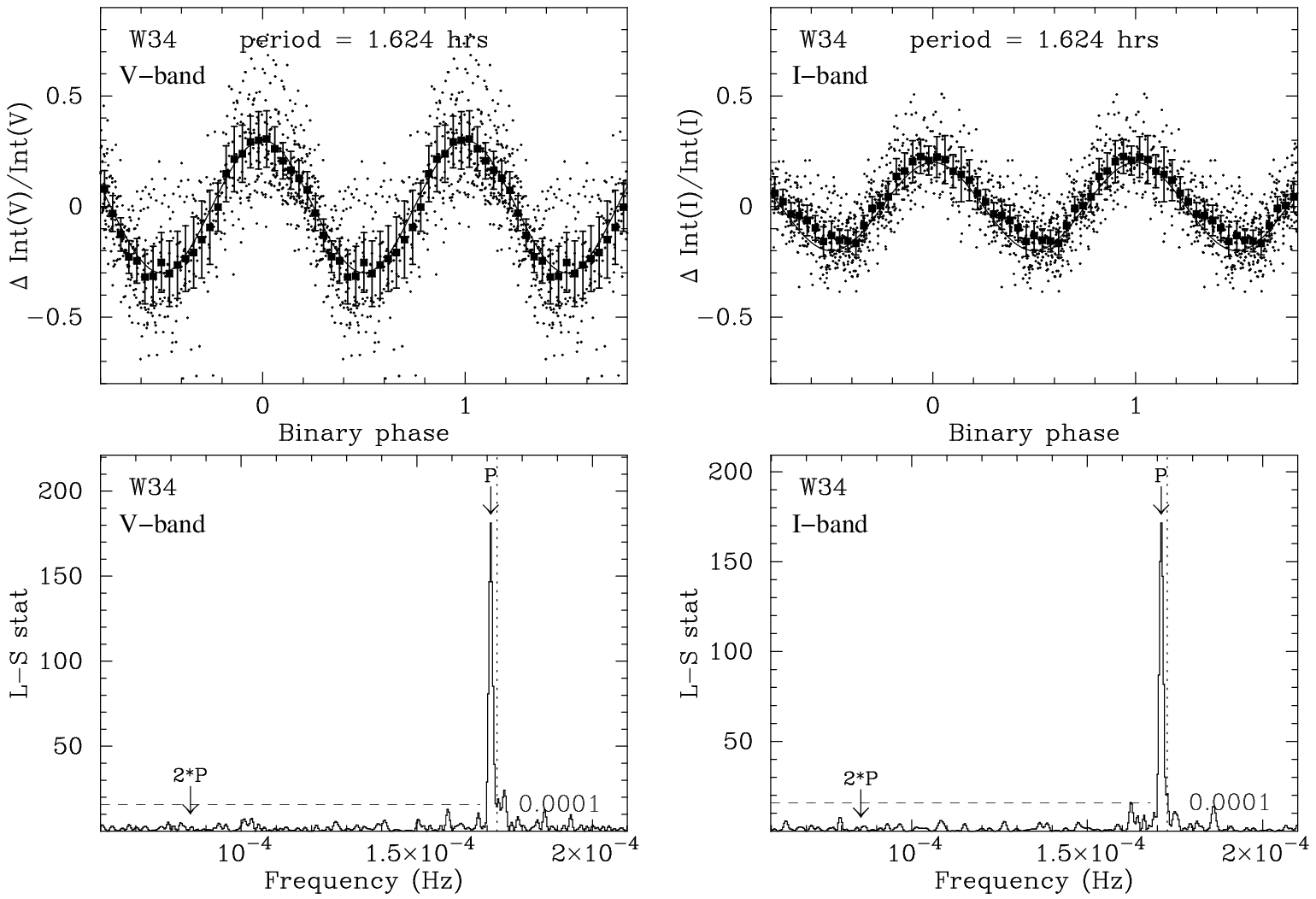,width=13.5cm}
\vspace*{0.0cm}
\caption{Phase plots and power spectra for \other. As explained in the text
the observed period is close to, but statistically different from, the \hst\
orbital period.}
\vspace*{-0.4cm}
\label{fig.w34phase}
\end{figure*}

\subsubsection{Color changes}
\label{sec.colchange}

We plot in Figure \ref{fig.w29col} the difference between the $V$ and
$I$-band relative intensity plots (see Fig. \ref{fig.w29phase}) as a
function of orbital phase. This figure is equivalent to a $V-I$ color vs
phase plot except that increasing y-values correspond to bluer colors
(unlike with magnitudes). Note that ({\em i}) \wopt\ clearly has a bluer
color at \phb\ $\sim$0.0 than at other phases (as expected if the secondary
is heated by irradiation from a luminous disk or MSP; see
\S~\ref{sec.disc}), and ({\em ii}) the minimum is at \phb\ $\sim$0.4,
rather than at \phb\ = 0.5. The main contribution to this possible light
curve asymmetry (or excess of $V$-band signal at \phb\ $\sim$0.5) is
visible in the $V$-band plot of Figure \ref{fig.w29phase}, where several
consecutive points lie above the sinusoidal model near \phb\ =
0.5. However, given the faintness of \wopt\ at these phases, it is
difficult to confidently declare that this is a real effect.

\subsection{Short Period Variable \other}

From a review of the optical variables we note that one other star has
properties similar to \wopt. A faint ($V$=22.3) star with blue colors ($U-V
\sim 0$) shows relatively large amplitude sinusoidal variations in both $V$
and $I$ and is only 4\farcs9 away from the cluster center.  The fit to the
GO-8267 time series of this star gives amplitude = 0.301(7), $P_b =
0.0676705(78)$ d, and $T_{max}$ = 51366.5573(3) MJD.  The likely \cha\
counterpart, W34 (Grindlay et al. 2001a), lies 0\farcs19 (2~$\sigma$) away
from the optical variable. It has a similar X-ray luminosity to that of W29
(see Table 2), and is also a moderately hard X-ray source, though
marginally softer than W29.

Two portions of the time series for \other\ are shown in Figure
\ref{fig.w34tseries}. The large deviations from the sinusoidal model will
be discussed in \S~\ref{sec.disc}.  The phase plots and power spectra for
\other\ are shown in Figure \ref{fig.w34phase} and the color vs phase plot
in Figure \ref{fig.w34col}. Note from the sinusoidal fit (and
Fig. \ref{fig.w34tseries}) that the period of \other\ (97.45 min) is close
to the \hst\ orbital period (96.4 min). However, we are convinced that the
variations of \other\ are real because: ({\em i}) the \other\ and \hst\
orbital periods differ at almost the 100~$\sigma$ level, ({\em ii}) in
cases where artifacts of the \hst\ orbit are present, there is usually
significant power at the harmonics of the \hst\ period and at
$\pm$(1/86400) Hz, but for \other\ the time series are very clean at the
one frequency, ({\em iii}) since the beat period between 97.45 min and 96.4
min (6.2 d) is shorter than the total observing window (8.3 d), there is a
full phase rotation of one cycle in the \other\ signal with respect to the
orbit of \hst, and ({\em iv}) the mean difference image (difference image
at \phb\ $\sim$0.0 minus the difference image at \phb\ $\sim$0.5) has a
stellar shaped PSF, which is only expected if the variation is real, rather
than an artifact from a bad pixel interacting with the \hst\ orbit-induced
motions to generate a false signal.

\begin{figurehere}
\vspace*{0.3cm}
\hspace*{0.1cm}
\epsfig{file=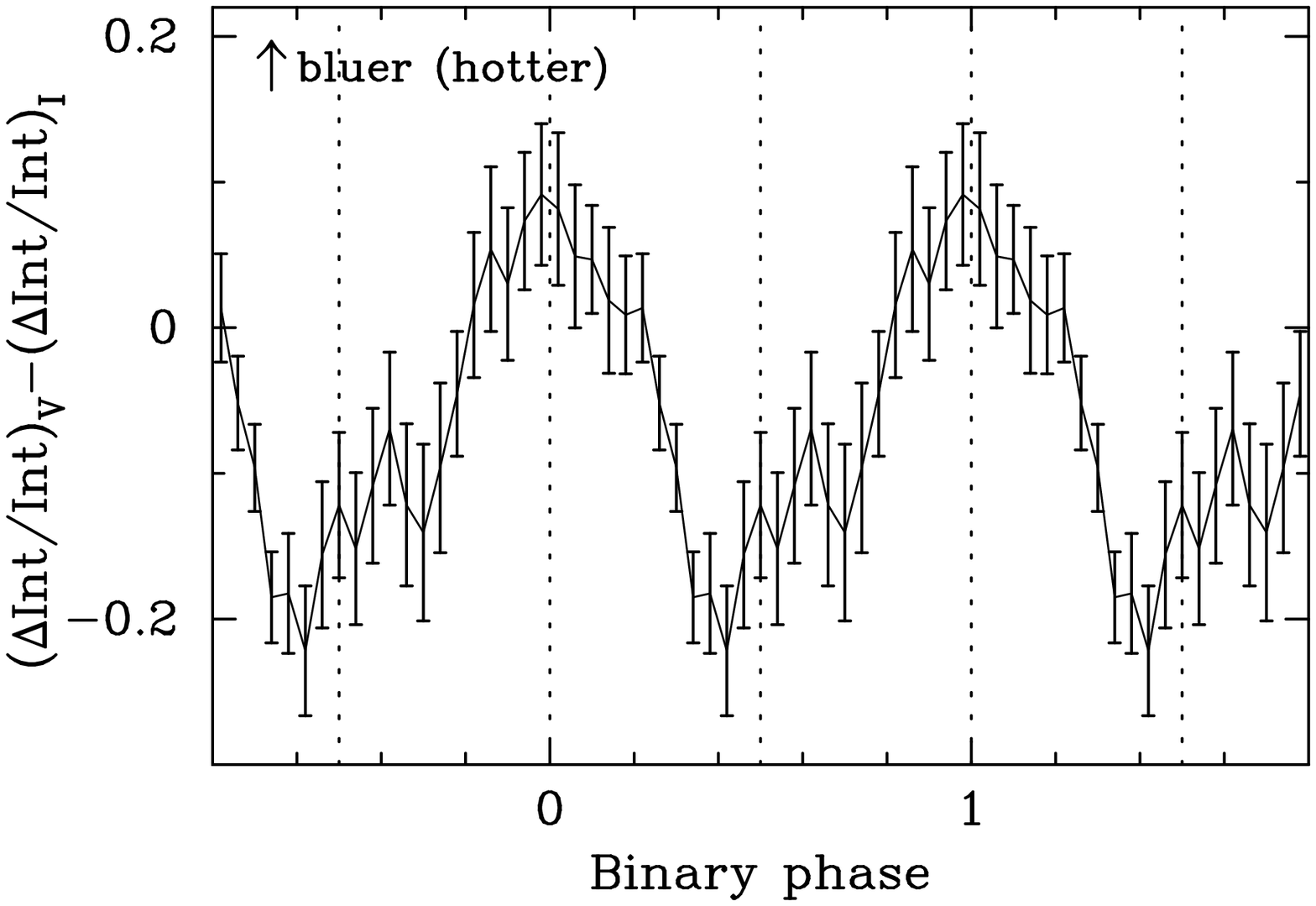,width=7.0cm}
\vspace*{0.0cm}
\caption{The difference between the $V$ and $I$ time series (from GO-8267)
plotted as a function of phase for \other.}
\vspace*{0.5cm}
\label{fig.w34col}
\end{figurehere}

\section{Discussion}
\label{sec.disc}

\subsection{\wopt}

As shown above, there is excellent period and phase agreement between
\wopt\ and 47 Tuc W and we regard the association between these two objects
as secure.  This marks the first time a radio pulsar has been localized
with sub-arcsecond precision by means other than radio pulse timing or
synthesis observations.  We now discuss the nature of \wopt\ and its
variability in the context of all that is known about the binary system,
including radio (\S~\ref{sec.radio}) and X-ray data.

As noted in \S~\ref{sec.radio}, the MSP 47 Tuc W was eclipsed for
$\sim$30\% of its orbit on the only day in which it was detected at radio
wavelengths.  The two eclipsing MSPs known in the field (PSRs B1957+20 and
J2051$-$0827, with companion masses of 0.02--0.03 \mdot) show, to first
order, persistent eclipses at similar orbital phases (other eclipsing MSPs
with similar binary parameters, in 47 Tuc, are far more poorly studied and
relatively little is known in detail about their eclipsing properties).  By
contrast, the MSPs in NGC~6397 (NGC 6397 A; D'Amico et al. 2001) and Terzan~5
(PSR~B1744$-$24A; Nice \& Thorsett 1992) show eclipses that are variable in
time and orbital phase.  These two systems, while significantly different
from each other, are in addition distinct from the other eclipsing MSPs in
having larger companion masses, $\sim$0.45 \mdot\ and $\sim$0.10 \mdot,
respectively.

Naturally we do not know if the eclipses in 47 Tuc W are persistent, but
the minimum companion mass of 0.13 \mdot\ suggests it is a fundamentally
different star from the $\sim$0.02 \mdot\ objects.  The companion mass is
possibly similar to that of 47 Tuc U's companion and other presumed He WD
companions to MSPs, but the observed eclipse could not be caused by a He WD
companion, and in any case as seen in Figure \ref{fig.cmds} the photometry
for \wopt\ appears inconsistent with a He WD.  The companion mass {\em is}
above the hydrogen burning limit.  It therefore appears plausible that
\wopt\ is an MS star, the first such known example as a companion to an MSP
(if the companion in NGC 6397 A is an evolved subgiant rather than an MS star).

The large amplitude, nearly sinusoidal light curve of \wopt\ is reminiscent
of the optical light curves of PSRs B1957+20 (Callanan et al. 1995) and
J2051$-$0827 (Stappers et al. 2001). This similarity suggests that the
optical variations of \wopt\ are caused by rotational modulation, as we see
alternately the heated and (relatively?) unheated sides of the tidally
locked companion once per orbit (as expected, \wopt\ appears to be hottest
when it is brightest; Fig. \ref{fig.w29col}). A similar, large amplitude
light curve may also be present in the companion to PSR~B1744$-$24A in
Terzan~5, but the formidable reddening and crowding in this cluster (Cohn
et al. 2002) have thwarted a recent attempt to detect the companion with
\hst\ (Edmonds et al. 2001b).

Since the blue color of \wopt\ probably results from the effects of
heating, we have estimated the effects and intensity of the `radiation
bath' \wopt\ experiences, based on the size of the binary and the expected
spin-down luminosity (\edot) of 47 Tuc W. Unfortunately, without a timing
solution, no direct estimate of \edot\ is possible. Using the observed
X-ray luminosity of 7.8$\times 10^{30}$ erg s$^{-1}$ and the \lx\ vs \edot\
relationship of Grindlay et al. (2002), we estimate that \edot\ for 47 Tuc
W is $\sim 3\times 10^{35}$ erg s$^{-1}$ (the 1~$\sigma$ range is $\sim
2-8\times 10^{35}$ erg s$^{-1}$). A much smaller value of \edot\
(7.8$\times 10^{33}$ erg s$^{-1}$) is derived using the linear relation of
Becker \& Tr\"umper (1999), \lx$\sim$10$^{-3}$\edot, derived largely for
field MSPs.

The stellar evolution models of Bergbusch \& Vandenberg (1992) give an
approximate guide to the relationship between luminosity and mass for a MS
star.  We adopt the lowest mass model given by Bergbusch \& Vandenberg
(1992), a 0.15 \mdot\ model with \teff\ $=3300$~K and $M_V=13.2$, and test
whether significant heating of such a star is expected.  We use the
measured orbital period, an assumed mass for the NS of 1.4 \mdot, and the
above secondary mass to estimate the binary separation from Kepler's Third
Law.  Then, using the estimated secondary radius from the stellar models,
and assuming that the MSP radiates its wind isotropically, the `high'
\edot\ estimate given above implies that the MSP energy intercepted by
\wopt\ should be $\sim 1 \times10^{33}$erg s$^{-1}$, a factor of $\sim 120
\times$ the luminosity of the assumed companion (by comparison, the
corresponding numbers for the companion to 47 Tuc U are $\sim7
\times10^{30}$erg s$^{-1}$ or 0.013$\times$ the luminosity of the
companion; EGH01).  Assuming that all of the spin-down energy is reradiated
as a blackbody, we derive a temperature of 15,500~K. This would imply that
heating effects are important and the MS star should be brighter and hotter
than in its unperturbed state, possibly explaining the blue colors and
relative brightness of \wopt.

The above temperature is an upper limit (unless there is beamed emission),
since a significant fraction of \edot\ could be powering other processes
such as mass loss. For example, for PSR 1957+20, Callanan et al. (1995)
find that the fraction of the MSP flux intercepted by the secondary that is
converted into optical emission (defined as $\eta$) is either in the range
0.07--0.2 for a Roche lobe filling model, or $\eta \gtrsim$ 3 (significant
beaming) for a secondary considerably underfilling its Roche lobe.
Stappers et al. (2001) find $\eta \sim$0.3 for J2051$-$0827 and EGH01 find
$\eta \sim 1$ for 47 Tuc U.  Guided by these numbers we assume a
conservatively small value for $\eta$ of $\sim$0.1 to give a reasonable
lower limit on the heating flux, and derive a blackbody temperature of
$\sim$8,700~K.  Therefore heating effects should still be important.
 
For this model, the ratio between the heating flux and the undisturbed
luminosity of the assumed stellar companion (120) is only a factor of 2.4
larger than the ratio (caused by heating) between the $V$-band flux of
\wopt\ and the $V$-band flux of the stellar model (implying $\eta
\sim$0.4).  However, if this model is approximately correct, the relatively
small difference in intensity between the heated and unheated sides of
\wopt\ (a factor of 6 compared to the above factor of 120) has to be
explained.  One possibility is that heating of the companion has occurred
fairly evenly over its surface. Although the secondary is probably tidally
locked, significant heating of the darker side of the star might be
expected because of energy transport within the star from the combined
effects of convection (this low mass MS star should be fully convective),
and the effects of relativistic $e^{+}e^{-}$ pairs and $\gamma$-ray photons
(D'Antona 1996). It is also possible that a crescent of the heated
hemisphere is visible at all orbital phases if the orbital inclination of
the binary is significantly less than 60\degr\ (Fruchter et al. 1995). This
requires a higher mass companion for consistency with the radio results and
the brighter star also gives a correspondingly smaller amplitude. In the
case of NGC 6397 A, where the inclination is known to be high, the
companion also has a much smaller variability amplitude than expected based
on heating estimates (Orosz \& van Kerkwijk 2002). This may be a feature
common to all MSP systems containing non-degenerate companions.

Approximate upper limits to the mass and radius of the 47 Tuc W companion
are set by the requirement that mass transfer be avoided.  Using the
Roche-lobe formula from \citet{pac71}
($r/a=\mathrm{0.462[(M_{W}/(M_{NS}+M_W)]^{1/3}}$, where $r$ is the
Roche-lobe radius, $a$ is the binary separation, $\mathrm{M_{W}}$ is the
mass of the secondary and $\mathrm{M_{NS}}$ is the mass of the NS), we
derive that, in its unperturbed state, a 0.29 \mdot\ model with \teff
$=3680$~K and $M_V=10.8$ would underfill its Roche lobe by $\sim$1\% (the
corresponding upper limit on the orbital inclination is 27\degr).  Our 0.15
\mdot\ model would (in its unperturbed state) significantly underfill its
Roche lobe (the ratio of the stellar radius to $r$ is 0.6), explaining why
the NS is not accreting.  However, under the influence of the extreme
radiation bath from the MSP, \wopt\ may expand to fill or almost fill its
Roche lobe (D'Antona \& Ergma 1993), with accretion being inhibited by
radiation pressure from the MSP (as may be occuring in NGC 6397 A; Burderi
et al. 2002).

Using the `low' \edot\ estimate given above, the MSP energy intercepted by
\wopt\ is only $\sim4 \times10^{31}$erg s$^{-1}$ (with a reradiated
blackbody temperature of 6,300~K), a factor of $\sim$3.3 times the
luminosity of the assumed companion. This luminosity ratio is smaller than
the observed variation and is a factor of 15 {\em smaller} than the ratio
between the $V$-band flux of \wopt\ and the $V$-band flux of the stellar
model. So, unless significant beaming is present, only a fainter MS star
than assumed above can give a larger ratio between the heating flux and the
stellar luminosity. However, the disparity between the $V$-band fluxes then
increases by the same amount.  A smaller ratio between the $V$-band fluxes
is given by a brighter, more massive secondary (requiring an inclination
$<$ 60\degr), but then the heating flux would be too small to cause the
observed variations.  These inconsistencies suggest that the higher \edot\
value given earlier is closer to the true value, providing indirect support
for the \lx\ vs \edot\ relationship of Grindlay et al. (2002).

The similarities between the X-ray properties of W29 and those of the X-ray
counterparts of NGC 6397 A (see \S~\ref{sec.ast}) and 47 Tuc J offer valuable
insight.  These similarities at the very least allow for a similar physical
origin of the X-ray emission.  The extended nature of the hard X-ray source
in NGC 6397 A suggests it is due to shocked gas lifted from the binary
companion (Grindlay et al. 2002), and similar behavior may be occurring in
W29. This match in X-ray properties between W29 and two other eclipsing
MSPs suggests evidence for a trend, since apart from 47 Tuc J, all 9 of the
X-ray-detected MSPs in 47 Tuc that are bright enough to have useful
hardness ratios determined have soft spectra (caused by likely thermal
emission from their polar caps; Grindlay et al. 2002).


As we have already mentioned, 47 Tuc W is an extremely unusual (and so far
possibly unique) MSP in having a likely MS companion. This likely
represents evidence for stellar interactions, not surprising for a system
so close to the cluster center. One possible formation scenario for this
system is that an NS was spun up by accretion to form an MSP, which
eventually evolved into an MSP/very low mass companion binary. Later, a
collision between the MSP binary and the currently observed MS star caused
the very low mass star (the lightest object amongst the 3 stars) to be
ejected from the system. Alternatively, a single MSP could have exchanged
into a double MS star binary, although in this case tight constraints on
the mass of the ejected star may be necessary unless \wopt\ was once
considerably more massive than it is today. Significant mass loss could be
taking place if \wopt\ is close to filling its Roche lobe and material is
being expelled from the system under the propeller mechanism (Burderi et
al. 2002). A third possibility is that the observed binary was formed as a
result of direct 2-body tidal capture (Mardling 1995) between a single MSP
and an MS star, although this process may be much less important than
binary interactions (Rasio et al. 2000).

In testing these various exchange scenarios, we note that 47 Tuc W is
located very close to the cluster center (only 3\farcs8 or 0.16$r_c$ away),
consistent with the centrally concentrated spatial distribution of the
other MSPs. If the MSP was initially ejected from the cluster core in an
exchange encounter it may have had time to sink back into the cluster core,
since the MS star could have been captured several Gyr ago, and the central
relaxation time for 47 Tuc is 0.1 Gyr (increasing to 3 Gyr at the half-mass
radius of 2\farcm8; Harris 1996). The NGC 6397 MSP, by contrast, is found
$\sim$30$''$ from the cluster center (Ferraro et al. 2001) and is therefore
well outside the cluster core ($r_c = 3 $\farcs0; Harris 1996), possibly
showing evidence for ejection from the core (D'Amico et al. 2001; Grindlay
et al. 2002). This ejection may have occurred quite recently, since the
central relaxation time for the core-collapsed NGC 6397 is only 0.08 Myr,
increasing to 0.3 Gyr at the half-mass radius of 2\farcm3 (Harris 1996). If
these numbers are accurate, a system as heavy as NGC 6397 A should sink
rapidly back into the core (unless it is in a highly elliptical orbit). One
possible explanation is that NGC 6397 A was formed very recently by an NS
capturing a subgiant, and was ejected from the core in this interaction.
This scenario is broadly consistent with the relatively small
(characteristic) age for the system of 0.35 Gyr (D'Amico et al. 2001). It
may also have been ejected from the cluster core in a subsequent
interaction. A similar ejection scenario may apply to PSR~B1744$-$24A,
which is located $\sim40''$ from the center of Terzan 5 (Nice \& Thorsett
1992), equivalent to $\sim5 r_c$ using the recent $r_c$ determination by
Cohn et al. (2002).

It may be possible that \wopt\ will eventually fill its Roche lobe and
cause mass accretion onto the NS to recommence, as may occur for NGC 6397 A
(Ferraro et al. 2001 and Burderi et al. 2002). In this scenario, as \wopt\
keeps losing mass, it eventually may end up like the very low mass
(0.02--0.03 \mdot) systems such as 47 Tuc J (though these may be ablated He
WDs rather than ablated MS stars; Rasio et al. 2000). Then, after continued
ablation, it could possibly become an isolated MSP.

\subsection{\other}

Unlike with \wopt, there is no period or phase match for \other\ with a known
MSP (its period is similar to, but statistically different from, the 95.39
$\pm$ 0.1 min period of 47 Tuc R, an eclipsing MSP with a very low mass
companion; CLF00), but a significant fraction of the 47 Tuc MSPs have not
yet been detected in the radio (CLF00; Grindlay et al. 2002).  Given the
short period of this system, the heating effects (if it is an MSP) may be
even more dramatic than for \wopt, since the ratio between the heating flux
and the stellar luminosity would be higher, assuming a similar luminosity
companion star. We note that the color difference between the brighter and
darker sides of the companion ($\sim$30\%) compared to the amplitude of the
variation is larger than the similar ratio for \wopt. The smaller amplitude
of \other\ compared to \wopt\ may simply be an inclination effect or it may
reflect a significant difference in the level of heating of the dark side
of the companion, or show evidence for a different type of star, such as a
very low mass degenerate star instead of an MS star.  We also note that in
this system the low mass model assumed above for \wopt\ would overfill its
Roche lobe by a few percent.

There are broad similarities in the appearance of the color vs phase plots
for \wopt\ (Fig. \ref{fig.w29col}) and \other\ (Fig. \ref{fig.w34col}),
particularly in the positions of the peak near \phb\ = 0.0, the minimum at
\phb\ $\sim$0.4, and the possible `hump' of relatively blue light at \phb\
= 0.4--0.7. The relatively blue color at optical maximum is secure, but the
effects near \phb\ = 0.5 can only be considered marginal, given the
faintness of the stars at these times and the possibility that artifacts of
the \hst\ orbit may have leaked into the time series (half the period of
\wopt\ is 95.72 min).

These comparisons do not constitute proof that \other\ is an MSP companion,
and therefore we briefly consider alternative explanations for its
behavior.  The near-sinusoidal light curve of \other, with its short period
and large amplitude, is similar to that of a W~UMa variable (a contact, or
near contact binary consisting of two MS stars). However, this
interpretation is clearly inconsistent with the blue colors of this object.
A second possible explanation is that \other\ is a CV, since most of the
CVs in 47 Tuc (and also most of the CVs in NGC 6397 and the field) are
optically variable blue stars with moderately hard X-ray counterparts
(Grindlay et al. 2001a). However, the likely CVs in 47 Tuc show flickering
(random fluctuations on timescales of minutes with amplitudes of
$\sim$0.05--0.1 mag) which hides the presence of periodic variations in
most cases. Of the few CVs in 47 Tuc which do show periodic variations, the
light curves are very different from those of \other\ (and \wopt), and show
either low amplitude ($\lesssim$ 0.1 mag) periodic variations from
ellipsoidal modulation, or a combination of ellipsoidal modulation and
eclipses (Edmonds et al. 2002a, in preparation), with distinctly
non-sinusoidal light curves (there is significant signal in the power at
both the orbital period and half the orbital period). Given these
differences, then unless \other\ is a different type of CV from the
$\sim$20 good candidates found in 47 Tuc, we consider it more likely to be
an MSP.

As shown in Figure \ref{fig.w34tseries}, there are several instances of
large deviations from the sinusoidal model for \other. We examined the
standard deviation of the residuals (data$-$sinusoid) for each cycle of the
\other\ light curve, after iteratively removing 3~$\sigma$ deviations from
the mean, and compared these results with those found for \wopt. The mean
standard deviation for \other\ (0.15) was similar to the value found for
\wopt\ (0.21), but there were some differences in the distributions. No
3~$\sigma$ deviations were found for \wopt, but two were found for \other\
(at HJD $\sim$51363.6 and HJD $\sim$51367.8), and two consecutive cycles of
$\sim 2.9 \sigma$ deviations (in HJD = 51370.35--51370.5) were also
found. These variations could be evidence for variable mass loss from the
companion or they could be episodes of unusually large flickering, if the
CV model is correct.

\section{Summary and Future Observations}

Briefly summarizing the key results in this paper, we have identified a
binary companion, \wopt, to a second MSP in the cluster 47 Tuc
(adding to the detection of a He WD companion to MSP 47 Tuc U; EGH01). This
faint, blue star shows large amplitude sinusoidal variations in both $V$
and $I$ (caused by heating of a tidally locked companion), and
has a period and phase which match remarkably well with those of the MSP 47
Tuc W.  We argue, based on the radio and optical results, that \wopt\ is a
main sequence star, a rare circumstance for an MSP companion, and we
present evidence for another optical variable (with similar properties to
\wopt) that may also be an MSP companion.

Deeper \cha\ observations of 47 Tuc scheduled for 2002 September (PI
J. Grindlay) will allow variability and spectral studies of W29, with
dependences on phase, to be made. Simultaneous observations scheduled with
ACS (as part of the same program) will provide the first \hst\ images of 47
Tuc in $H_{\alpha}$, and should detect evidence for a strong emission line
in \other\ if it is a CV rather than an MSP companion. Combination of this
data with archival data and with the GO-8267 time series will refine the
period and phase of \other.  Finally, by combining these X-ray data with
existing and future \hst\ observations of 47 Tuc, more MSP companions may
be identified.

\acknowledgments

We thank Andy Fruchter and Kailash Sahu for discussions, and the referee
Scott Ransom for helpful comments.  This work was supported in part by
STScI grant GO-8267.01-97A (PDE and RLG) and NASA grants NAG~5-9095 and SAO
grant GO1-2063X (FC).



\tabletypesize{\footnotesize}

\begin{deluxetable}{lllll}
\tablecolumns{5}
\tablewidth{0pc}
\tablecaption{Summary of optical data}
\tablehead{\colhead{Dataset} &
 \colhead{Observation Date} & \colhead{Camera} & 
 \colhead{Filters} & \colhead{Exposures}
}
\startdata


GO-8267      & 1999 July 3--11 & WFPC2 & F555W (`$V$') & 636$\times$160~s \\
             &  & WFPC2 & F814W (`$I$')  & 653$\times$160~s \\
GO-7503      & 1999 October 28 & WFPC2 & F555W  & 9$\times$40~s;
             3$\times$120~s\tablenotemark{a} \\
GO-8219      & 1999 September 10 & STIS & CLEAR & 8$\times$30~s \\
             & 2000 August 7--16 & STIS & CLEAR & 38$\times$30~s \\
SM3/ACS-9028 & 2002 April 5 & ACS/HRC & F475W  & 20$\times$120~s \\

\tablenotetext{a}{The 40~s exposures were composed of co-added 20~s exposures}


\enddata
\end{deluxetable}




\tabletypesize{\tiny}

\begin{deluxetable}{lcccccllcc}
\tablecolumns{10}
\tablewidth{0pc}
\tablecaption{Optical, radio and X-ray data for the
companion to MSP 47 Tuc W (\wopt) and a candidate MSP
counterpart (\other)}
\tablehead{\colhead{Source} &
 \colhead{RA\tablenotemark{a}} & \colhead{Dec\tablenotemark{a}} & 
 \colhead{$U$} & \colhead{$V$} & \colhead{$I$} & \colhead{$P_b$} &
 \colhead{$T_{max}$\tablenotemark{b}} & \colhead{\lx\tablenotemark{c}} & 
 \colhead{X-ray spectrum\tablenotemark{c}} \\
\colhead{} & \colhead{(J2000)} & \colhead{(J2000)} & \colhead{} & \colhead{} & \colhead{} & 
\colhead{(days)} & \colhead{(MJD)} & \colhead{(erg s$^{-1}$)} & \colhead{(photon index)} 
}
\startdata


\wopt\ & 00 24 06.07(1) & $-$72 04 49.02(6) & 23.8(5) & 22.3(2) & 21.6(3) &
0.132944(1)\tablenotemark{d} & 51366.8705(3) & 7.8$\times10^{30}$ & 
1.8$\pm$0.6 \\
 & & & & & & & 51214.3837(12)\tablenotemark{e} & & \\
47 Tuc W & \nodata & \nodata  & \nodata  & \nodata  & \nodata  & 0.13295(4)\tablenotemark{f} & 51214.38454(3)\tablenotemark{g} & \nodata & \nodata \\
\other\ & 00 24 05.21(2) &  $-$72 04 46.59(6) & 22.3(5) & 22.3(3) & 21.6(4)  &
 0.0676705(78)  & 51366.5573(3) & 4$\times10^{30}$ & $2.4^{+0.5}_{-0.6}$ \\

\tablenotetext{a}{Coordinates are in the JPL DE200 planetary ephemeris frame used
for MSP timing (Freire et al. 2001a)}
\tablenotetext{b}{Time of optical maximum (see text)}
\tablenotetext{c}{\lx\ and the spectral fit assume cluster $N_H = 3.0\times10^{20}$cm$^{-2}$}
\tablenotetext{d}{Binary period derived from optical data}
\tablenotetext{e}{Optical $T_{max} - 1147$ orbits}
\tablenotetext{f}{Binary period derived from radio data}
\tablenotetext{g}{$T_{max}$ derived from radio data}


\enddata
\end{deluxetable}




\end{document}